\documentclass{article}
\usepackage{spconf,amsmath,graphicx,hyperref}
\usepackage{algorithm}
\usepackage{algorithmic}
\usepackage{pifont}
\usepackage{newfloat}
\usepackage{listings}
\usepackage{amsmath}
\usepackage{booktabs} %
\usepackage{amssymb}

\title{CoSP: Reconfigurable Multi-State Metamaterial Inverse Design via Contrastive Pretrained Large Language Model}
\name{
    Shujie Yang$^{\star \dagger \ddagger}$ \qquad 
    Xuzhe Zhao$^{\star \dagger \ddagger}$ \qquad 
    Yuqi Zhang$^{\star \dagger \ddagger}$ \qquad
    Yansong Tang\thanks{Corresponding authors: tang.yansong@sz.tsinghua.edu.cn;}$^{\star}$ \qquad 
    Kaichen Dong\thanks{dkc22@sz.tsinghua.edu.cn}$^{\star \dagger \ddagger}$ 
    \thanks{This work has been submitted to the IEEE for possible publication. Copyright may be transferred without notice, after which this version may no longer be accessible.}
}

\address{
    $^{\star}$ Institute of Data and Information, Tsinghua Shenzhen International Graduate School,\\ Tsinghua University, Shenzhen, Guangdong 518055, China \\
    $^{\dagger}$ Center of Double Helix, Tsinghua Shenzhen International Graduate School, \\ Tsinghua University, Shenzhen, Guangdong 518055, China \\
    $^{\ddagger}$ Intelligent Passive Thermal Control Center, Research Institute of Tsinghua University in Shenzhen, \\ Shenzhen, Guangdong, 518057, China
}

\begin{document}
\ninept
\maketitle
\begin{abstract}
Metamaterials, known for their ability to manipulate light at subwavelength scales, face significant design challenges due to their complex and sophisticated structures. Consequently, deep learning has emerged as a powerful tool to streamline their design process. Reconfigurable multi-state metamaterials (RMMs) with adjustable parameters can switch their optical characteristics between different states upon external stimulation, leading to numerous applications. However, existing deep learning-based inverse design methods fall short in considering reconfigurability with multi-state switching. To address this challenge, we propose \textbf{CoSP}, an intelligent inverse design method based on contrastive pretrained large language model (LLM). By performing contrastive pretraining on multi-state spectrum, a well-trained spectrum encoder capable of understanding the spectrum is obtained, and it subsequently interacts with a pretrained LLM. This approach allows the model to preserve its linguistic capabilities while also comprehending Maxwell's Equations, enabling it to describe material structures with target optical properties in natural language. Our experiments demonstrate that CoSP can design corresponding thin-film metamaterial structures for arbitrary multi-state, multi-band optical responses, showing great potentials in the intelligent design of RMMs for versatile applications.
\end{abstract}

\section{Introduction}
\label{sec:intro}
\begin{figure}[t]
    \centering
    \includegraphics[width=0.8\linewidth]{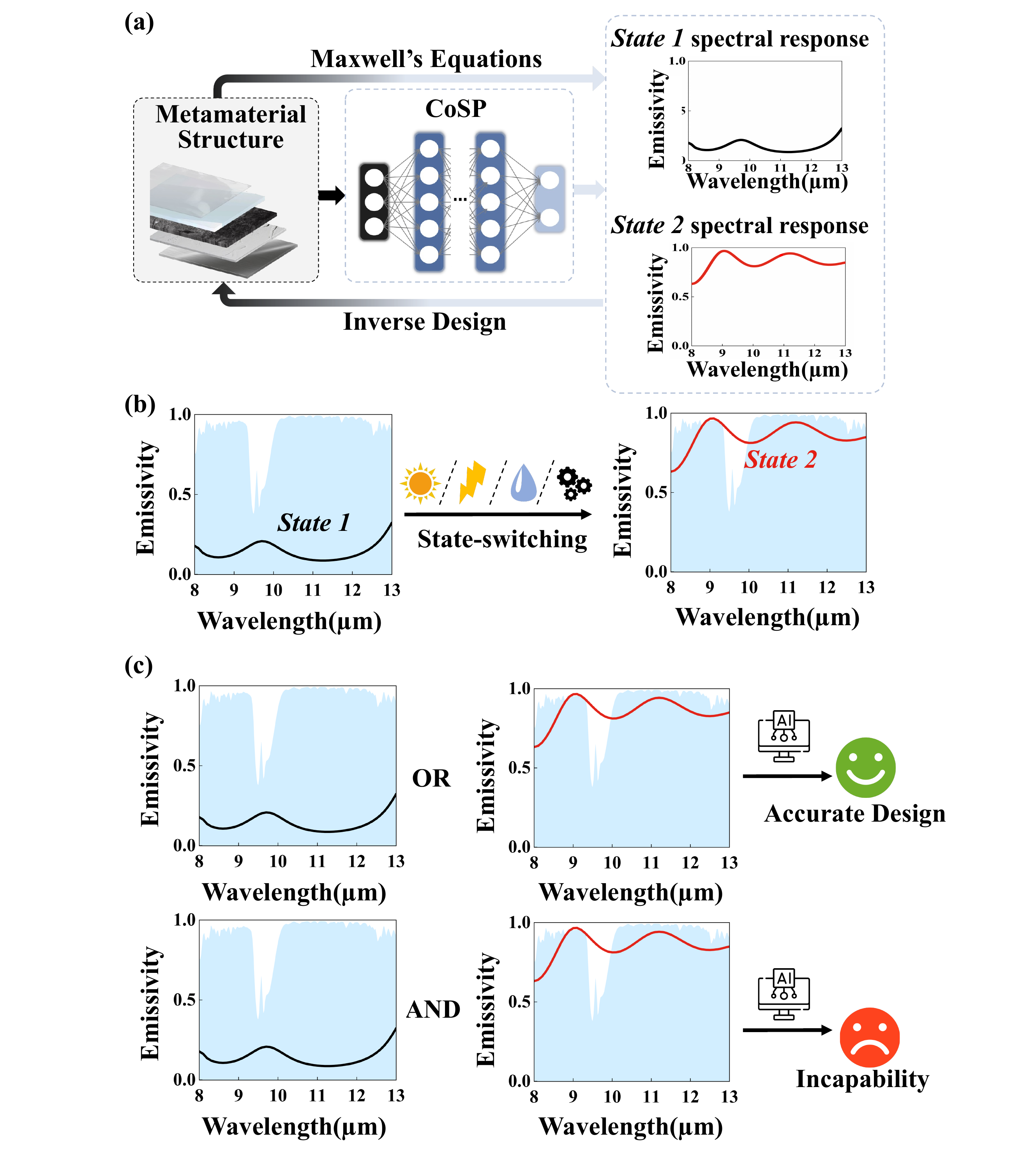}
    \caption{Schematic diagrams illustrating \textbf{(a)} the concept of inverse design, where metamaterial structures are designed based on desired physical responses, \textbf{(b)} the concept of RMMs, which exhibit distinct changes in physical properties (thermal emissivity in this example) before and after state switching, and \textbf{(c)} the limitation of existing inverse design methods that can only aim at one single state, rather than accommodating multiple states.}
    \label{fig:schematic}
\end{figure}
Metamaterials, composed of individual `meta-atoms', enable the design of materials with extraordinary properties through precise engineering the microarchitectures \cite{smith2004metamaterials}. They exhibit properties scarce or impossible in natural materials, impacting industries like information science \cite{zhang2018space}, optoelectronics \cite{zheludev2012metamaterials}, biomedical sciences \cite{sreekanth2016extreme}, and energy \cite{raman2014passive}. However, designing metamaterials by adjusting meta-atom sizes and arrangements via trial and error is time-consuming and difficult, and this experience-driven approach lacks transferability between disciplines. Recently, AI-for-Science with deep learning has shown great promise for metamaterials design and optimization \cite{ma2021deep}.

Inverse design for photonic metamaterials aims to find optimal material compositions and structures for desired optical properties like absorption, reflection, and transmission. Traditional methods such as FDTD and FEM simulations heavily rely on exhaustive parameter sweeps, making them inefficient in high-dimensional design spaces. AI has emerged as a tool for data-driven design, learning relationships between structures and optical responses. Approaches like Bayesian optimization, deep learning, and reinforcement learning accelerate discovery by navigating design spaces. However, existing AI-based inverse design methods still face limitations: low design freedom, restricted to fixed materials and predefined structures \cite{unni2020deep,chen2023broadband}, and poor transferability, requiring retraining for new objectives or materials \cite{sakurai2019ultranarrow,yu2023general,xi2023ultrahigh}. Moreover, no effective algorithm exists for inverse design of reconfigurable multi-state metamaterials (RMMs) with distinct optical states \cite{wang2021automated,ma2024optogpt}, limiting development of adaptive metamaterials with dynamic functionality.

Reconfigurable Multi-state metamaterials refer to metamaterials that have multiple operational states, driven by factors such as electricity, heat, humidity, and mechanical forces, and can meet the needs for state switching in different scenarios. Figure \ref{fig:schematic} illustrates some fundamental concepts about multi-state metamaterials and the shortcomings of previous inverse design methods. However, multi-state metamaterials exhibit intelligent and dynamic self-regulation of their physical properties and hold a very promising application prospect. 

In this paper, we propose \textbf{CoSP}, a RMM inverse design method via contrastive pretrained large language model. This method employs contrastive learning to study the optical responses of metamaterials across multiple states (for example, optical absorptivity and reflectivity). CoSP first maps these optical responses to a unified latent spectral space and then interacts it with an LLM, retaining linguistic capabilities while precisely designing metamaterials with target optical responses. Through a variety of experiments, CoSP's inverse design capabilities to design RMMs from hypothetical spectra have been robustly validated. In summary, CoSP represents a significant advancement in the field of intelligent RMMs design, offering a novel approach to address the challenges posed by multi-state physical systems.

\begin{figure}[t]
    \centering
    \includegraphics[width=\columnwidth]{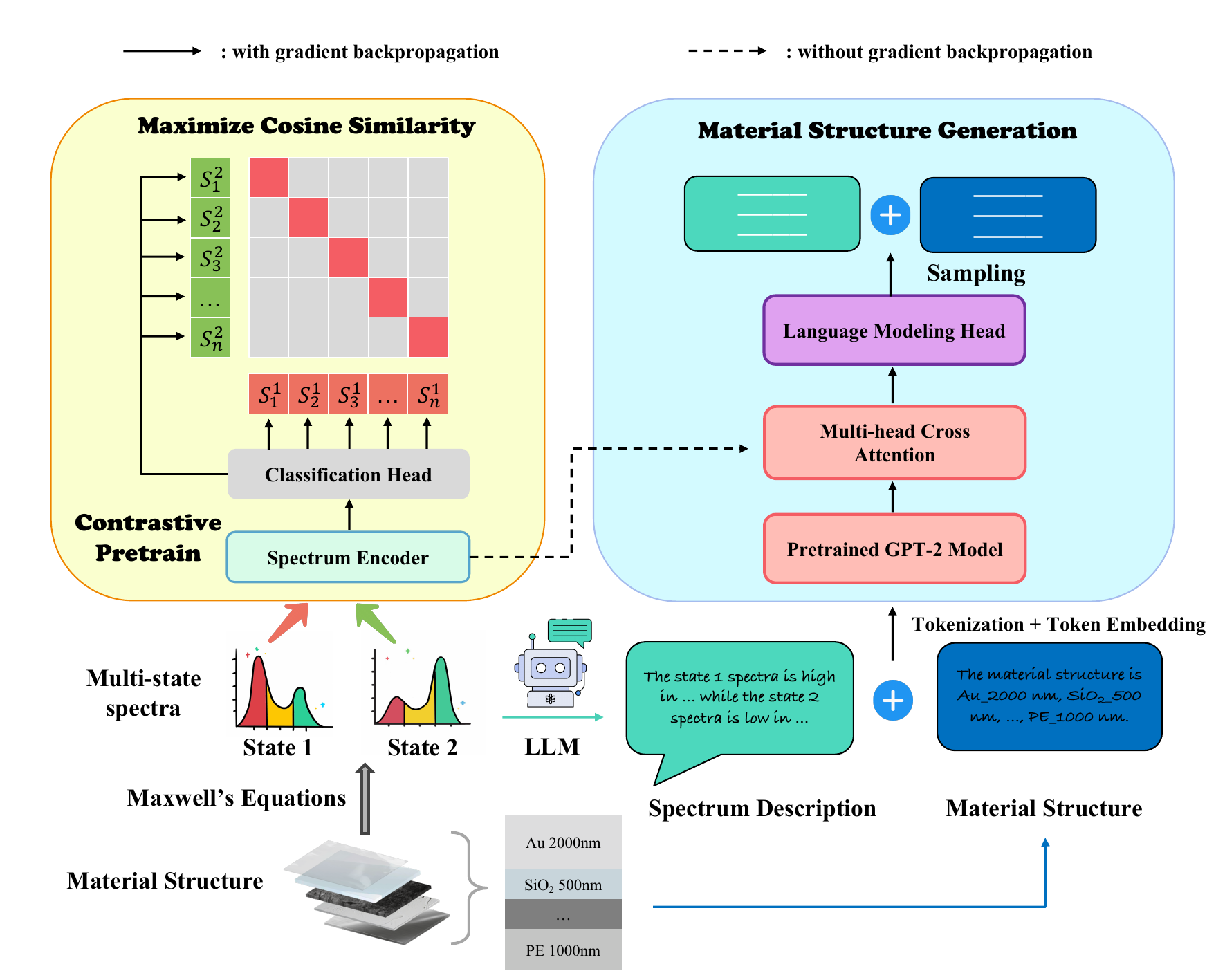}
    \caption{Architecture of the proposed CoSP framework for generating RMMs with multi-state optical responses in free text sequence.}
    \label{fig:copp}
\end{figure}

\section{Methodology}
\label{sec:methodology}

As introduced above, RMMs can exhibit adjustable optical responses, including optical absorptivity, reflectivity, and transmissivity. Designing such multi-state metamaterials is a challenging task for existing inverse design methods, as they are unable to simultaneously address distinct design targets of the same metamaterial. To tackle this challenge, we propose \textbf{CoSP} (\underline{co}ntrastive multi-\underline{s}tate \underline{p}retraining). The architecture of CoSP is shown in Figure \ref{fig:copp}, which consists of two stages: contrastive multi-state spectrum pretraining and metamaterial structure generation with a pretrained spectrum encoder.

\subsection{Contrastive Multi-State Spectrum Pretraining}
We utilize a Transformer encoder \cite{transformer} as the spectrum encoder backbone, which includes components such as wavelength embedding, self-attention, and a feed-forward network. Notably, unlike the text encoder used in CLIP \cite{clip}, token embedding is not used in this configuration. This is based on our understanding that text is discrete in the lexical space, while spectra are continuous across bands. Token embedding could disrupt the continuity of spectra, leading the model to favor a more segmented interpretation of the input, which is not ideal for spectral data.

Initially, the spectrum $\mathbf{S}^j_i$ of different states $\mathbf{j}$ of the same metamaterial $\mathbf{i}$ is projected by a linear layer $Proj$, and wavelength encoding $WavEncoding$ is subsequently applied to the projected spectrum representation. In the absence of token embedding, the wavelength encoding is an absolute positional encoding applied in the sequence direction. Subsequently, the spectrum representation is processed through the $N$-layer transformer encoder blocks $H$, yielding the output.
The encoder blocks are followed by a fully connected layer $ClsHead$, the classification head, used solely during the contrastive pretraining stage. According to SimCLR \cite{simclr}, using a nonlinear fully connected layer (i.e., one that includes an activation function) leads to more effective representations than a linear fully connected layer. This process can be formally described by:
\begin{equation}
    \begin{aligned}
    \mathbf{s}^j_i &= Proj(\mathbf{S}^j_i)\\
    \mathbf{s}^j_i &= \mathbf{s}^j_i + WavEncoding(\mathbf{s}^j_i)\\
    \mathbf{s}^j_i &= ClsHead(H(\mathbf{s}^j_i))\\
    \end{aligned}
\end{equation}

Then this approach employs contrastive learning to maximize the similarity of spectrum representations $\mathbf{s}^j_i$ of different states $j$ of the same metamaterial $i$ through the InfoNCE (Information Noise Contrastive Estimation) loss:
\begin{equation}
\begin{aligned}
L_j=-\frac{1}{n}\sum_{i=1}^n\log&\frac{\exp(\cos(\mathbf{s}_i^j,\mathbf{s}_i^{j^{*}})/\tau)}{\sum_{k=1}^n\exp(\cos(\mathbf{s}_i^j,\mathbf{s}_k^{j^{*}})/\tau)} \\
L=&\frac{1}{2} (L_1 + L_2)
\end{aligned}
\end{equation}
where $n$ is the batch size, $j^{*}$ denotes another state distinct from $j$, $\tau$ is a learnable temperature parameter to control the distribution of the similarity matrix.Thereby a spectrum encoder is trained in a self-supervised manner.

\subsection{Spectrum Description}
Due to the one-to-many mapping from optical spectra to metamaterial designs, different metamaterial designs may exhibit similar optical spectra. As a result, the training targets of inverse design are not limited to material sequences alone. Otherwise, the model is confined to the metamaterial design space within the training set, thereby losing its generalization capability for new but similar spectra. Translating target spectra into human-readable descriptions can significantly improve the generalization of the inverse design method as all similar spectra can be recognized by the model to avoid the retraining process. Moreover, by observing the spectra in natural language described by CoSP, the model's true understanding of the spectra can be perceived and verified by the user, that is, whether the spectrum representation learned by the spectrum encoder is effective. 

Hence, unlike existing inverse design methods that solely and directly generate metamaterial designs, CoSP is preceded by a spectrum description, which is a depiction of the target multi-state spectra generated by an LLM. It encompasses the local maxima and minima (i.e., peaks and troughs) of the spectra across various states and wavelength bands. The detailed generation process and data examples will be introduced later in the experimental results section.

\subsection{Inverse Design - Material Structure Generation}
We employed the Transformer decoder architecture to generate sequences of thin film metamaterial structures with targeted optical responses. We initialized the main components of the decoder—namely, the text embedding matrix, self-attention mechanism, and language modeling head—with the pretrained weights of GPT-2 \cite{gpt2}. By doing so, we leveraged GPT-2's ability to understand the underlying textual semantics. We fed the spectrum representation, obtained from an encoder pretrained through spectrum contrast, into the multi-head cross-attention module within the Transformer decoder:
\begin{equation}
\begin{aligned}
\mathbf{Z} &= \text{Softmax}\left(\frac{\mathbf{Q}\mathbf{K}^{\top}}{\sqrt{d_k}}\right) \mathbf{V} \\
\mathbf{Q} &= \mathbf{W}_q \mathbf{q}, \quad \mathbf{K} = \mathbf{W}_k \mathbf{s}, \quad \mathbf{V} = \mathbf{W}_v \mathbf{s}
\end{aligned}
\label{eq:cross_attention_qkv}
\end{equation}

where \(\mathbf{Q}\) represents the query matrix, derived by applying the weight matrix \(\mathbf{W}_q\) to the text embedding \(\mathbf{q}\). The key matrix \(\mathbf{K}\) and the value matrix \(\mathbf{V}\) are obtained by applying the weight matrices \(\mathbf{W}_k\) and \(\mathbf{W}_v\) to the spectrum representation \(\mathbf{s}\), respectively. This interaction enabled the model to effectively incorporate the context of the spectrum representation, contributing to the generation of coherent, meaningful, and physically compliant material descriptions. The interaction between the spectral representation derived from the encoder and the decoder allows the model to comprehend the relationship between materials and their spectra across different phases and wavelength bands; in other words, the model is essentially understanding Maxwell's equations. During the training phase, we employed Causal Language Modeling (CLM) as the training objective to optimize our model. Causal Language Modeling involves training the model to predict the next token in a sequence, given the preceding tokens. 
\begin{equation}
\mathcal{L}_{\text{CLM}} = -\sum_{i=1}^{n} \log P(x_i \mid x_1, x_2, \ldots, x_{i-1})
\label{eq:clm}
\end{equation}

where \(\mathcal{L}_{\text{CLM}}\) is the loss function for Causal Language Modeling. This objective involves predicting each token \(x_i\) in a sequence based on the preceding tokens \(x_1, x_2, \ldots, x_{i-1}\), thereby ensuring that text generation occurs in a causal, forward manner. Precisely, this unidirectional autoregressive generation method aligns remarkably with the multi-layered structure of metamaterials.

\section{Experimental Results}
\label{sec:experimental results}

\subsection{Preliminary Setup}

In this section, RMMs integrated with vanadium dioxide (VO$_2$) are selected as an example to demonstrate the inverse design of multi-state target spectrum using CoSP. VO$_2$ \cite{tang2021temperature, wang2021scalable, guo2024durable} is a widely used phase-change material that undergoes a reversible, solid-solid phase transition at 68°C, transforming from an insulating phase (\textbf{I-phase}, corresponding to \textbf{State I} of RMMs) to a metallic phase (\textbf{M-phase}, corresponding to \textbf{State M} of RMMs) \cite{fu2013comprehensive}.

\subsection{Main Results}
\subsubsection{Inverse Design}
In this section, three samples will be selected from the test set to demonstrate the inverse design. For one sample, the absorptivity spectrum pair in two states are input as targets, which also serve as the inputs. Subsequently, CoSP will generate spectrum descriptions and corresponding material structures. We then employ Transfer Matrix Method (TMM) \cite{song2023transfer} to calculate the multi-state absorptivity of these RMMs and compare them with the target absorptivity spectrum pair. To illustrate the dissimilarity between the selected samples and the training set, we also present the “most accurate” spectrum in the training set that is closest to the target spectrum in terms of Mean Squared Error (MSE). The MSE of a spectrum pair is calculated by first computing the MSE of each individual spectra against the target, and then summing the MSE of the two spectra to obtain the final MSE. We both calculated the MSE between the designed RMM spectrum and the target spectrum (hereby coined as MSE$\_$designed), as well as the MSE between the “most accurate” spectrum in the training set and the target spectrum (hereby coined as MSE$\_$closest). Figure \ref{fig:results} displays the results of the inverse design. From these results, we can draw the following conclusions:
\begin{figure}[h]
    \centering
    \includegraphics[width=\columnwidth]{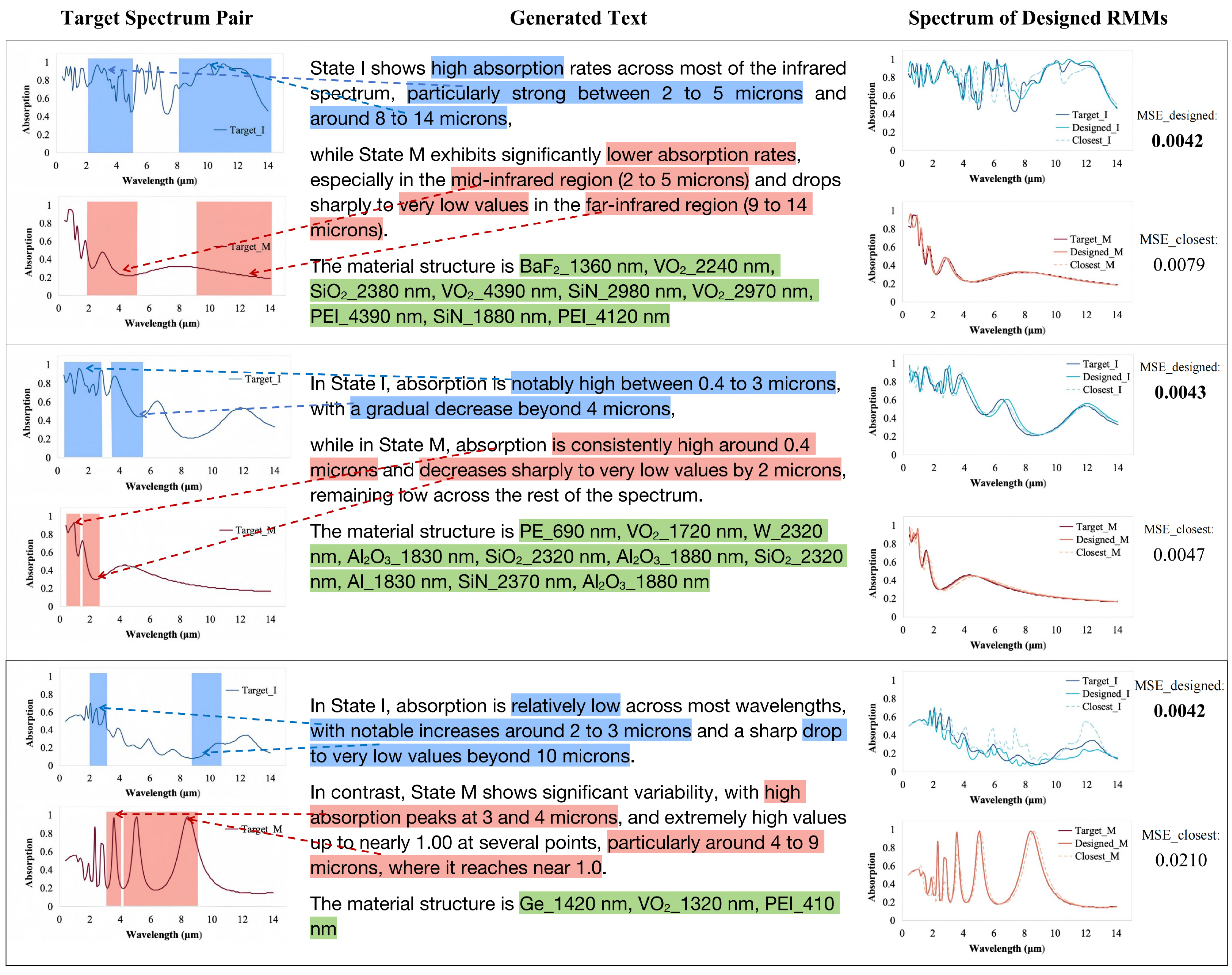}
    \caption{CoSP inverse design of RMMs for general purposes. \textbf{Target Spectrum Pair} represents the input spectrum with two states. \textbf{Generated Text} comprises the natural language results produced by CoSP, including the spectrum description and material structure sequence of the designed RMM. The spectrum description, highlighted in color, corresponds to the wavelength bands depicted in the first column's spectra. \textbf{Spectrum of Designed RMMs} illustrates the spectrum calculated using TMM for the material structure generated by CoSP. Additionally, the sample in the training set with the closest match to the target spectrum is displayed, along with the MSE calculated for both, to demonstrate the generalization capability of CoSP.}
    \label{fig:results}
\end{figure}
\begin{itemize}
    \item The spectrum descriptions generated by CoSP accurately depict the characteristics of the input spectra in different states, thereby strongly demonstrating that the Spectrum Encoder has learned highly useful and high-quality spectral representations and features. It also validates the effectiveness of the Contrastive Multi-State Spectrum Pretrain.
    \item Numerical calculations reveal that the material structures generated by CoSP possess spectrum pairs that are essentially consistent with the target spectrum pair simultaneously, perfectly accomplishing the inverse design of RMMs.
    \item By comparing MSE$\_$designed and MSE$\_$closest, it can be observed that CoSP has designed RMMs with performance overwhelming the whole training set, reflecting the strong generalization capability of CoSP.
\end{itemize}

\subsubsection{Exploration of hypothetical spectrum}
In the practical design of metamaterials, it is often desired that the optical response of these materials exhibits extreme characteristics within critical wavelength bands. For instance, a high absorptivity in the visible wavelength band, a low absorptivity in the near-infrared (IR) wavelength spectrum, and again a high absorptivity in the mid-IR wavelength spectrum. Such complicated requirements lead to a target optical spectrum that mimics a square wave pattern. However, in practical terms, no material (i.e., neither natural material nor artificial metamaterials) exhibits such a square-wave spectral profile. To demonstrate the potential of CoSP in real-world applications, we aim to show its capability to explore such square-wave-like hypothetical spectra that are otherwise too difficult to design. %

We select infrared (IR) camouflage \cite{tang2020thermal} as the design objective for this exploration. IR camouflage is a technique that achieves camouflage from infrared detectors by designing an RMM coating that adjusts the multi-spectral emissivity and temperature of an object's surface \cite{zhu2021multispectral}. 
In this application, three IR spectral bands are involved: infrared detectors typically operate in the 2-5 $\mu m$ and 8-13 $\mu m$ bands \cite{xi2023ultrahigh}, while the 5-8 $\mu m$ band, which is undetectable by IR sensors, is used for radiative cooling to help regulate temperature \cite{qin2023whole}. Therefore, at high environmental temperatures (\textbf{State M}), the RMM coating should have \textbf{low emissivity in the 2-5 $\mu m$ and 8-14 $\mu m$ bands} as well as \textbf{high emissivity in the 5-8 $\mu m$ band} for strong radiative cooling, all of which helps reduce thermal radiance from the RMM coating. Conversely, at low environmental temperatures \textbf{(State I)}, the RMM coating should \textbf{have high emissivity in the 2-5 $\mu m$ and 8-14 $\mu m$ bands} as well as \textbf{low emissivity in the 5-8 $\mu m$ band} for heat-retaining, leading to enhanced thermal radiance from the RMM coating. %
\begin{figure}[h]
    \centering
    \includegraphics[width=\columnwidth]{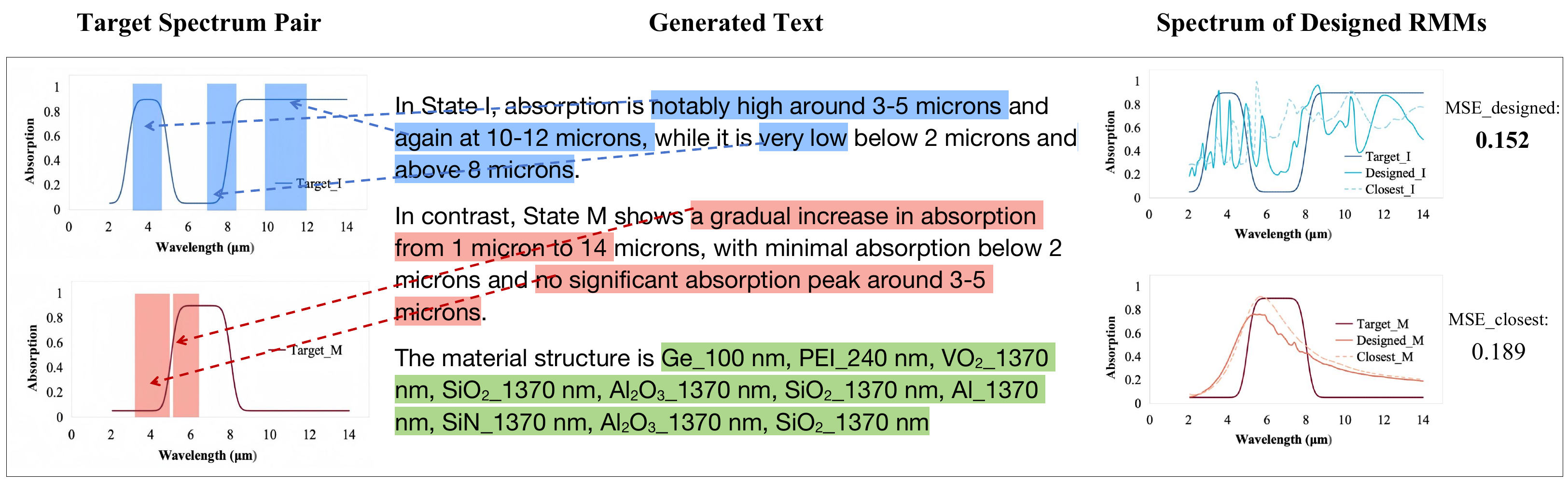}
    \caption{Illustration of inverse design results aimed at IR-camouflage.}
    \label{fig:camouflage}
\end{figure}

In the experiment, we set the wavelength-dependent emissivity of the target spectrum pair to 0.9 or 0.05 according to the design target of IR camouflage. We then apply a Gaussian filter to smooth the square-shaped waveforms for better inverse design performance before inputting it into CoSP as the design target. Figure \ref{fig:camouflage} displays the results for IR-camouflage applications. It is evident that, despite the target spectrum pair representing an ideal scenario that is otherwise impossible in nature, CoSP is capable of designing RMMs that closely approximate the target while maintaining accurate spectral descriptions. Moreover, the designed materials outperform the best samples in the training set, further demonstrating CoSP's robust generalization capability. In summary, CoSP effectively bridges the gap between ideal and real-world scenarios in metamaterial design.

\begin{figure}[htbp]
    \centering
    \includegraphics[width=0.7\columnwidth]{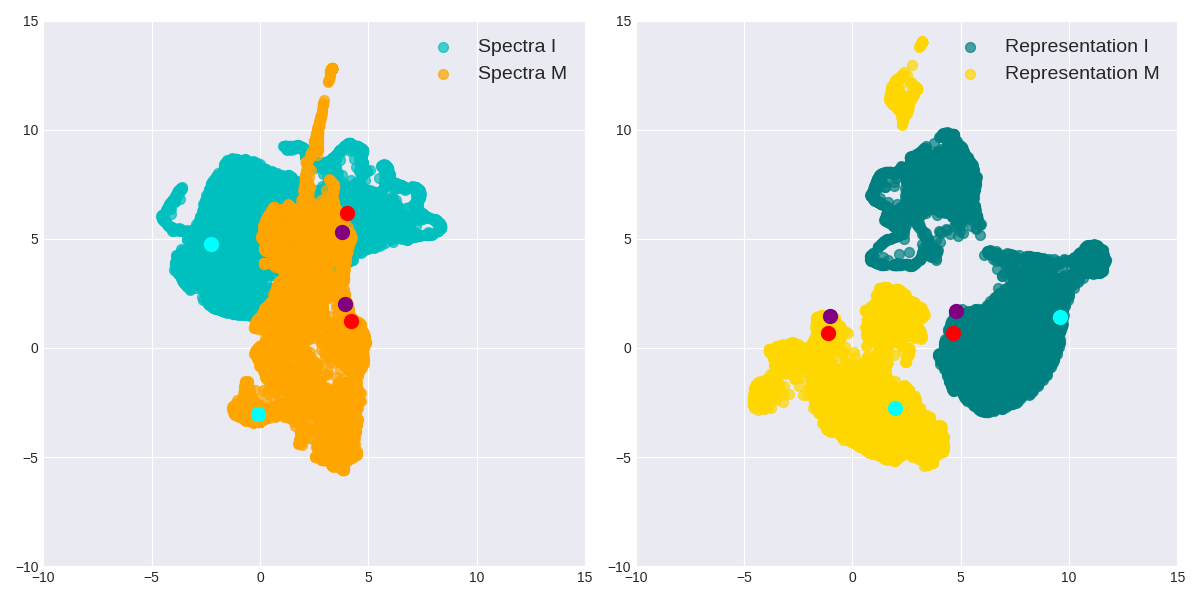}
    \caption{Visualization of raw spectrum (left) and spectrum representation (right) after UMAP dimensionality reduction. Three pairs of representative spectrum pairs are highlighted as point pairs with different colors (red, purple, and blue).}
    \label{fig:umap}
\end{figure}
\subsection{Spectrum Representation Quality}
CoSP learns representations that distinguish states and match spectra from the same sample. UMAP visualization in Figure~\ref{fig:umap} shows that while raw spectra from State I and State M overlap, their learned representations form clearly separated clusters. The relative distances of three sample pairs (red, purple, light blue) are preserved between the raw and latent spaces, showing the encoder maintains data relationships while separating states.

\begin{figure}[h]
    \centering
    \includegraphics[width=0.7\columnwidth]{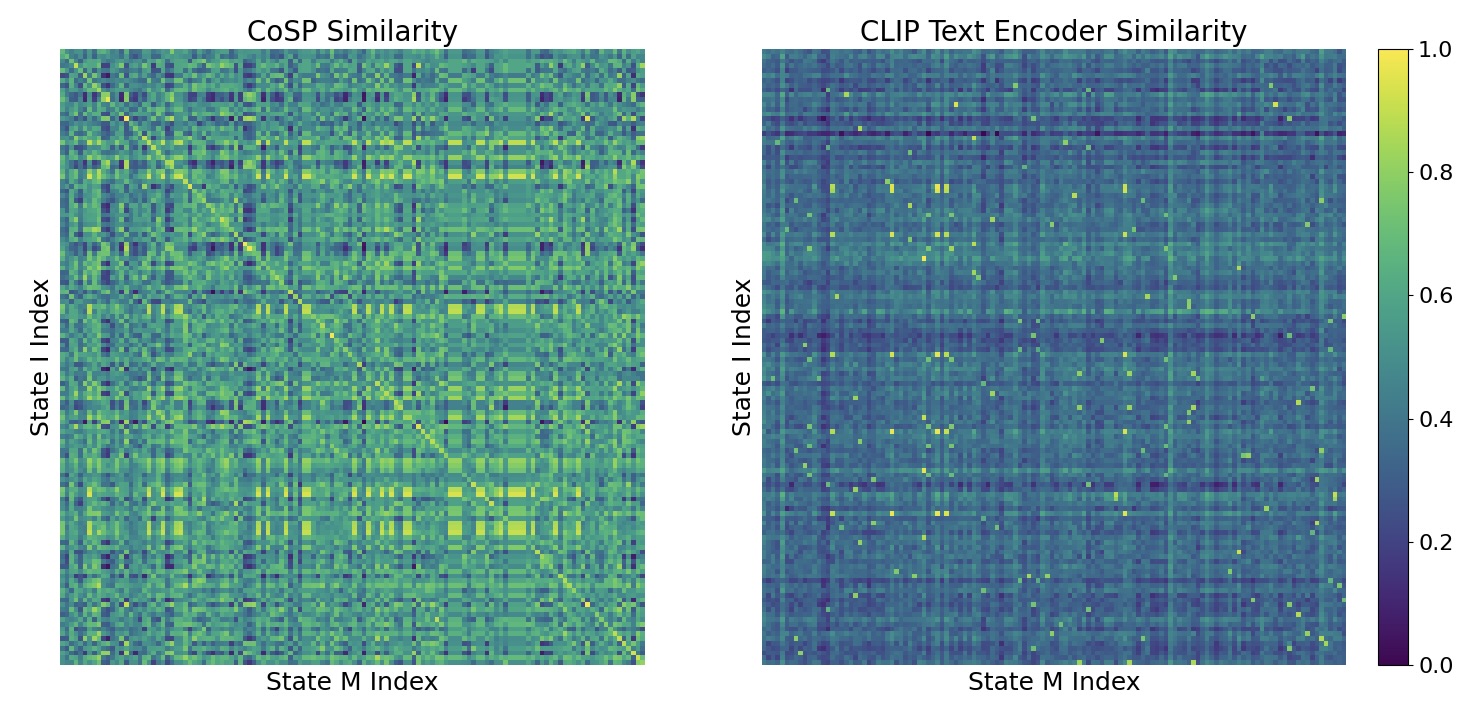}
    \caption{Heatmaps of the similarity matrices obtained using the CoSP spectrum encoder (left) and the CLIP text encoder (right).}
    \label{fig:heatmap}
\end{figure}

We compute a similarity matrix \(X = \mathbf{S}^I (\mathbf{S}^M)^\top\) from encoded test samples. As shown in Figure~\ref{fig:heatmap}, CoSP's matrix exhibits a bright diagonal, indicating high within-sample similarity between states, unlike CLIP. This result confirms CoSP produces discriminative and correspondence-aware representations, achieving its goal.

\section{Conclusion}
\label{sec:conclusion}

In this paper, we introduce \textbf{CoSP}, a novel inverse design model for multi-phase metamaterials. This model leverages contrastive multi-phase spectrum pretraining and large language models to achieve the inverse design of multi-phase metamaterials for the first time, encompassing all capabilities of previous models. Experimental results demonstrate that CoSP not only excels in designing spectra within the test set but also achieves precise approximation of spectra corresponding to hypothetical square waveforms that likely do not exist in the real world. This enables the exploration of unknown yet potentially useful spectrum, paving the way for the design of metamaterials with entirely new functions. The significance of this work lies in its groundbreaking contribution to the field, offering a transformative approach to the discovery and realization of advanced metamaterials.

\vfill\pagebreak

\bibliographystyle{IEEEbib}
\bibliography{refs}
\section{Acknowledgement}
S. Yang, X. Zhao, Y. Zhang, and K. Dong acknowledge support from National Natural Science Foundation of China (grant 62375151), Shenzhen Science and Technology Program (grant KJZD20240903095721028),  National Key R\&D Program of China (grant 2023YFB3208700), Start-up funding in Tsinghua Shenzhen International Graduate School, Tsinghua University.
\end{document}